\documentstyle[12pt]{article}
\begin{document}
\title{{\bf A sufficient and necessary condition for superdense coding of quantum states}}
\author{Daowen Qiu\\
\small{Department of Computer Science, Zhongshan University,
Guangzhou 510275,}\\{\small People's Republic of China}\\
{\small E-mail address: issqdw@mail.sysu.edu.cn}}
\date{  }

\maketitle
\begin{center}
\begin{minipage}{130mm}
\begin{center}{\bf Abstract}\end{center}
{\small  Recently, Harrow {\it et al.} [Phys. Rev. Lett. {\bf 92,}
187901  (2004)] gave a method for preparing an arbitrary quantum
state with high success probability by physically transmitting
some qubits, and by consuming a maximally entangled state,
together with exhausting some shared random bits. In this paper,
we discover that some states are impossible to be perfectly
prepared by Alice and Bob initially sharing those entangled states
that are superposed by the {\it ground states}, as the states to
be prepared. In particular, we present a sufficient and necessary
condition for the states being enabled to be exactly prepared with
probability one, in terms of the initial entangled states (maybe
nonmaximally) superposed by the ground states. In contrast, if the
initially shared entanglement is maximal, then the probabilities
for preparing these quantum states are smaller than one.
Furthermore, the lower bound on the probability for preparing some
states are derived.

}


\end{minipage}
\end{center}
\vskip 10mm

\section*{1. Introduction}

Entanglement is one of the most intriguing phenomena in quantum
information theory and plays a pivotal role in quantum information
processing [1,2], including superdense coding [3], quantum
teleportation [4], remote state preparation [5], quantum
algorithms [6], and quantum cryptograph [7]. Superdense coding,
originally introduced by Bennett and Wieser [3] is the surprising
utilization of entanglement to enhance the capacity of a quantum
communication channel. That is, by making use of shared
entanglement, it is possible to communicate two classical bits by
physically transmitting only one qubit [3]. In a more general
fashion, if one shares $\log_{2}d$ ebits of entanglement, then one
can extract $2\log_{2}d$ classical bits of information by sending
a $d$-level quantum system (a qudit). The relationship between
quantum teleportation and superdense coding was investigated by
Werner and the others [8].

To date, superdense coding has been generalized in different
manners (for example, see [9] and the references therein). There
are mainly two scenarios: one is concerning communication between
multiparties [10] and the other is regarding nonmaximally
entangled states initially shared by Alice and Bob [11-15].
Hausladen {\it et al.} [12] showed that if Alice and Bob share a
nonmaximally entangled state then the capacity of dense coding
scheme is not $2\log_{2}d$ but equal to $S(\rho_{AB})+\log_{2}d$
bits of information in the asymptotic limit, where $S(\rho_{AB})$
is the entropy of entanglement of the shared state, and satisfies
$0\leq S(\rho_{AB})\leq \log_{2}d$. Therefore, we cannot
deterministically send $2\log_{2}d$ bits using nonmaximally
entangled states. Indeed, Mozes {\it et al.} [14] have dealt with
the relationship between the entanglement of a given nonmaximally
entangled state and the maximum number of alphabets which can be
perfectly transmitted in a deterministic fashion. However, if the
scheme is allowed to work in a probabilistic manner, then it
should be possible to send $2\log_{2}d$ bits of information by
sharing a nonmaximally entangled state [15]. Furthermore, it was
shown that, by initially sharing some W-states [16], superdense
coding and teleportation can also be perfectly performed [17].

Recently, another scheme, called {\it superdense coding of quantum
states} was proposed by Harrow, Hayden, and Leung [18].
(Furthermore, Abeyesinghe {\it et al.} [19] dealt with preparing
entangled states with  minimal cost of entanglement and quantum
communication.) That is, if the sender knows the identity of the
state to be sent, then two qubits can be communicated with a
certain probability by physically transmitting one qubit and
consuming one bit of entanglement [18]. Superdense coding of
quantum states is analogous to remote state preparation [5] but
the classical communication is now replaced by quantum
communication. To be precise, the purpose of superdense coding of
quantum states is to prepare a quantum state in Bob's system or
``sharing" a state that is entangled between Alice and Bob's
systems, for which Alice and Bob initially share a maximally
entangled state, and Alice first performs a physically operation
on her party with a certain success probability and then sends it
to Bob. Furthermore, Harrow {\it et al.} [18] presented a protocol
succeeding with high probability for communicating a $2l$-qubit
quantum state but some shared random bits are necessarily
consumed, besides transmitting $l+o(l)$ qubits and consuming $l$
ebits of entanglement.

However, if the shared randomness is {\it not} required, Hayden,
Leung and Winter [20] proposed a different protocol of superdense
coding of quantum states that can always successfully perform the
physical process, but may not guarantee the result to be {\it
exact}. Rather, the protocol may result in an approximate outcome
with high fidelity.

A natural question is that if Alice and Bob initially share
nonmaximally entangled states then how about the success
probability for preparing a quantum state; or, to prepare a
quantum state, could we fix on an appropriately partially
entangled state firstly shared by Alice and Bob, leading to the
optimal success probability? As we know, due to the Schmidt
Decomposition Theorem [1], any bipartite quantum state
$|\psi\rangle=\frac{1}{\sqrt{d}}\sum_{i,j=1}^{d}x_{i,j}|i\rangle_{A}|j\rangle_{B}$
(where $\{|i\rangle_{A}|j\rangle_{B}\}_{1\leq i,j\leq d}$ is an
orthonormal basis for ${\bf C}^{d}\bigotimes{\bf C}^{d}$) can be
written in the form $\sum_{j}r_{j}|e_{j}\rangle|f_{j}\rangle$ with
$r_{j}\geq 0$, where $\{|e_{j}\rangle\}$ and $\{|f_{j}\rangle\}$
are two orthonormal bases of systems $A$ and $B$, respectively,
so, it is evident that, by sharing this state they can exactly
prepare this state $|\psi\rangle$ with probability one. As we
know, the ground states are in general easier to be prepared.
However, $\{|e_{j}\rangle\}$ and $\{|f_{j}\rangle\}$ may not equal
to the {\it ground states} $\{|i\rangle_{A}\}$ and
$\{|j\rangle_{B}\}$, respectively, so, we here ask that, if the
initial entangled states (maybe nonmaximally) are superposed by
the ground states (i.e., $\{|e_{j}\rangle\}=\{|j\rangle_{A}\}$ and
$\{|f_{j}\rangle\}=\{|j\rangle_{B}\}$), as the states to be
prepared, then how about the superdense coding of quantum states?
The main goal of this paper is to clarify this question in detail.

The remainder of the paper is structured as follows. In Section 2,
we recall an exact probabilistic protocol of superdense coding of
quantum states, in which Alice and Bob initially share a maximally
entangled state, and Alice implements a transformation on her
party with a certain success probability and then sends it to Bob.
Section 3 is the main part and we discover that some states are
impossible to be perfectly prepared by Alice and Bob initially
sharing those entangled states that are superposed by the ground
states, as the states to be prepared. In particular, we present a
sufficient and necessary condition for the states being enabled to
be exactly prepared with probability one. Furthermore, the lower
bound on the probability for preparing some states are derived.
Finally some remarks are made in Section 4.

\section*{2. A probabilistic protocol for preparing quantum
states}

In this section, we recall a probabilistic protocol for preparing
quantum states which was dealt with by Harrow {\it et al.} [18].

Suppose we want to prepare a $d^{2}$-dimensional state
$|\psi\rangle=\frac{1}{\sqrt{d}}\sum_{i,j=1}^{d}x_{i,j}|i\rangle_{A}|j\rangle_{B}$
in Bob's system, by sending $\log_{2}d$ qubits and consuming
$\log_{2}d$ ebits of shared entanglement, where
$\{|i\rangle_{A}|j\rangle_{B}\}_{1\leq i,j\leq d}$ is an
orthonormal basis for ${\bf C}^{d}\bigotimes{\bf C}^{d}$. The
procedure can be described as follows. Alice and Bob initially
share $\log_{2}d$ ebits, or equivalently the maximally entangled
state
\begin{equation}
|\Phi_{d}\rangle=\frac{1}{\sqrt{d}}\sum_{i=1}^{d}|i\rangle_{A}|i\rangle_{B}.
\end{equation}
Alice performs a physical operation $X$ on her party and then
sends it to Bob, which may result in the state $|\psi\rangle$ to
be prepared with a certain success probability. We can represent
it by Equation (2):
\begin{equation}
(X\otimes
I)|\Phi_{d}\rangle=|\psi\rangle=\frac{1}{\sqrt{d}}\sum_{i,j=1}^{d}x_{i,j}|i\rangle_{A}|j\rangle_{B},
\end{equation}
where $I$ denotes the identity operator.

Nevertheless, $X$ may not be unitary, so the above scheme for
successfully preparing fixed $|\psi\rangle$ depends on the
successful application of $X$. One method to carry out $X$ is by
the generalized measurement $\rho\rightarrow \sum_{k}E_{k}\rho
E_{k}^{\dag}$ with Kraus operators [21,22]:
\begin{equation}
E_{0}=\frac{X}{\|X\|},\hskip 5mm E_{1}=\sqrt{I-E_{0}^{\dag}E_{0}},
\end{equation}
where the operator norm $\|X\|$ of $X$, is taken to be the square
norm, i.e., the square root of the largest eigenvalue of
$X^{\dag}X$. When the measurement outcome is 0, $X$ is
successfully performed, and the success probability $P_{s}$ is
then
\begin{equation}
P_{s}=Tr
E_{0}^{\dag}E_{0}\frac{I}{d}=\frac{Tr(X^{\dag}X)}{d\|X^{\dag}\|\|X\|}.
\end{equation}
From Equations (1, 2) we know that
$X|j\rangle_{A}=\sum_{j=1}^{d}x_{i,j}|i\rangle_{A}$ for
$j=1,2,\ldots,d$. Therefore, we have
\begin{eqnarray*}
Tr(X^{\dag}X)&=&\sum_{j=1}^{d}\langle j|X^{\dag}X|j\rangle\\
&=&\sum_{j=1}^{d}\sum_{i_{1},i_{2}=1}^{d}x_{i_{1},j}^{*}x_{i_{2},j}\langle
i_{1}|i_{2}\rangle\\
&=&\sum_{j,i=1}^{d}|x_{i,j}|^{2}\\
&=&d
\end{eqnarray*}
where the last equality results from $\langle \psi|\psi\rangle=1$;
$\{|j\rangle: j=1,2,\ldots,d\}$ is an orthonormal basis of system
$A$, as above.

Due to $Tr(X^{\dag}X)=d$, and $\|X^{\dag}\|\|X\|=\|X^{\dag}X\|$,
we further have
\begin{equation}
P_{s}=\frac{1}{\|X^{\dag}X\|}.
\end{equation}
 Clearly, if
$|\psi\rangle$ is a maximally entangled state, i.e.,
$|\psi\rangle=\frac{1}{\sqrt{d}}\sum_{i=1}^{d}|i\rangle_{A}|i\rangle_{B}$,
then $P_{s}=1$; if
$|\psi\rangle=\frac{1}{\sqrt{d}}\sum_{i,j=1}^{d}x_{i,j}|i\rangle_{A}|j\rangle_{B}$
with $\sum_{i=1}^{d}x_{i,j_{1}}^{*}x_{i,j_{2}}=0$ for any
$j_{1}\not= j_{2}$, then $X^{\dag}X={\rm diag}
(a_{1},a_{2},\ldots,a_{d})$ where
$a_{i}=\sum_{j=1}^{d}|x_{j,i}|^{2}$, and, consequently,
\begin{equation}
P_{s}=\frac{1}{\max(a_{1},a_{2},\ldots,a_{d})}.
\end{equation}
From equation (6) it follows that when
$\max(a_{1},a_{2},\ldots,a_{d})>1$, $P_{s}<1$. Therefore, we
consider that it is possible to increase the probability $P_{s}$
by changing the  maximally entangled state $|\Phi_{d}\rangle$
initially shared by Alice and Bob. Indeed, we will show that, in
terms of the state $|\psi\rangle$ to be prepared, the state
$|\Phi_{d}\rangle$ shared by the two parties can be, instead,
partially entangled to lead to the success probability one. This
is in contrast to superdense coding, in which partially
(nonmaximally) entangled state will decrease the bits of
information  communicated between Alice and Bob [12,14].

\section*{3. Superdense coding of quantum states with partially
entangled states}

Motivated by the issue addressed above, in this section we try to
answer it by deriving  appropriate partially entangled states, for
preparing some fixed states  with perfect success.

Let $\{|i\rangle_{A}|j\rangle_{B}\}_{1\leq i,j\leq d}$ be a given
orthonormal basis for ${\bf C}^{d}\bigotimes{\bf C}^{d}$. Suppose
state
$|\psi\rangle=\frac{1}{\sqrt{d}}\sum_{i,j=1}^{d}x_{i,j}|i\rangle_{A}|j\rangle_{B}$
to be prepared, where \begin{equation}\sum_{i,j}|x_{i,j}|^{2}=d.
\end{equation}
By means of the Schmidt Decomposition Theorem [1], there are
orthonormal bases $\{|e_{i}\rangle_{A}:i=1,2,\ldots,d\}$ and
$\{|f_{i}\rangle_{B}:i=1,2,\ldots,d\}$ of systems $A$ and $B$,
respectively, such that
\begin{equation}
|\psi\rangle=\frac{1}{\sqrt{d}}\sum_{i,j=1}^{d}x_{i,j}|i\rangle_{A}|j\rangle_{B}=\sum_{i=1}^{d}\lambda_{i}|e_{i}\rangle_{A}|f_{i}\rangle_{B}
\end{equation}
for some $\lambda_{i}\geq 0$, $i=1,2,\ldots,d$, with
$\sum_{i=1}^{d}\lambda_{i}^{2}=1$. Therefore, if Alice and Bob
initially share state
$\sum_{i=1}^{d}\lambda_{i}|e_{i}\rangle_{A}|f_{i}\rangle_{B}$,
then they can clearly prepare the desired state $|\psi\rangle$
succeeding with probability one.

In general, the ground states are easier to be prepared.
Therefore, we naturally ask if the entangled states initially
shared by Alice and Bob are superposed by the ground states, as
the state $|\psi\rangle$ to be prepared, i.e., the initial
entangled states have the form
$\sum_{i=1}^{d}\mu_{i}|g_{i}\rangle_{A}|h_{i}\rangle_{B}$ with
$\sum_{i=1}^{d}|\mu_{i}|^{2}=1$, where
\begin{equation}
\{|g_{i}\rangle_{A}:i=1,2,\ldots,d\}=\{|i\rangle_{A}:i=1,2,\ldots,d\},
\end{equation}
\begin{equation}
\{|h_{i}\rangle_{B}:i=1,2,\ldots,d\}=\{|i\rangle_{B}:i=1,2,\ldots,d\},
\end{equation}
then whether  $|\psi\rangle$ can be exactly prepared with
probability one by sharing some appropriate states
$\sum_{i=1}^{d}\mu_{i}|g_{i}\rangle_{A}|h_{i}\rangle_{B}$
(equations (9,10) are required) between Alice and Bob? Now, we
give a proposition to verify that this may not be true.

{\it Proposition 1.} If the state
$|\psi\rangle=\frac{1}{d}\sum_{i,j=1}^{d}|i\rangle_{A}|j\rangle_{B}$
is to be prepared, then with any entangled state of the form
$\sum_{i=1}^{d}\mu_{i}|g_{i}\rangle_{A}|h_{i}\rangle_{B}$
(equations (9,10) are required) initially shared by Alice and Bob,
the protocol of superdense coding of quantum states described
above can {\it not} perfectly prepare $|\psi\rangle$.

{\it Proof.} We present a proof by contradiction. If the protocol
of superdense coding of quantum states could exactly prepare
$|\psi\rangle$ with probability one, then there exists unitary
transformation $U_{A}$ on system $A$ such that
\begin{equation}
(U_{A}\otimes
I)\sum_{i=1}^{d}\mu_{i}|g_{i}\rangle_{A}|h_{i}\rangle_{B}=|\psi\rangle=\frac{1}{d}\sum_{i,j=1}^{d}|i\rangle_{A}|j\rangle_{B}
\end{equation}
for some $\mu_{i}$ with $\sum_{i=1}^{d}|\mu_{i}|^{2}=1$ and
equations (9,10) holding. Suppose that
\begin{equation}
U_{A}|g_{i}\rangle_{A}=\sum_{j=1}^{d}a_{j,i}|j\rangle_{A}, \hskip
2mm i=1,2,\ldots,d.
\end{equation}
Then the unitarity of $U_{A}$ results in
\begin{equation}
\sum_{j=1}^{d}a_{j,i_{1}}a_{j,i_{2}}^{*}=\left\{\begin{array}{ll}
1,& i_{1}=i_{2},\\
0,& i_{1}\not=i_{2}.
\end{array}
\right.
\end{equation}
With equation (12) we have
\begin{equation}
(U_{A}\otimes
I)\sum_{i=1}^{d}\mu_{i}|g_{i}\rangle_{A}|h_{i}\rangle_{B}=\sum_{i,j=1}^{d}\mu_{i}a_{j,i}|j\rangle_{A}|h_{i}\rangle_{B},
\end{equation}
which together with equation (11) results in
\begin{equation}
\mu_{i}a_{j,i}=\frac{1}{\sqrt{d}},\hskip 2mm
i,j=1,2,\ldots,d.\end{equation} Thus,
\begin{equation}
a_{1,k}=a_{2,k}=\ldots=a_{d,k}=\frac{1}{\sqrt{d}}e^{i\theta_{k}}
\end{equation}
for some real numbers $\theta_{k}$, $k=1,2,\ldots,d$. Therefore,
for any $i_{1}\not=i_{2}$,
\begin{equation}
\sum_{j=1}^{d}a_{j,i_{1}}a_{j,i_{2}}^{*}=e^{i(\theta_{i_{1}}-\theta_{i_{2}})}\not=0,
\end{equation}
a contradiction to equation (13). The proposition has been
verified. \hfill $\Box$

As well,  Proposition 1 clearly implies that the state
$|\psi\rangle=\frac{1}{d}\sum_{i,j=1}^{d}|i\rangle_{A}|j\rangle_{B}$
can not be prepared with probability one by initially sharing the
maximally entangled state
$\frac{1}{\sqrt{d}}\sum_{i=1}^{d}|i\rangle_{A}|i\rangle_{B}$
between Alice and Bob.

However, if the state $|\psi\rangle$ to be prepared satisfies a
certain condition, we can still choose an appropriate initial
state having the same orthonormal vectors as those in
$|\psi\rangle$, and by sharing this state exactly prepare
$|\psi\rangle$. This is further described by the following
theorem.

{\it Theorem 1.} Let
$|\psi\rangle=\frac{1}{\sqrt{d}}\sum_{i,j=1}^{d}x_{i,j}|i\rangle_{A}|j\rangle_{B}$
be the state to be prepared. Then $|\psi\rangle$ can be exactly
prepared with probability one by initially sharing the entangled
state
$|\Phi_{d}\rangle=\sum_{i=1}^{d}c_{i}|i\rangle_{A}|i\rangle_{B}$
between Alice and Bob for some $c_{i}$ with
\begin{equation}
\sum_{i=1}^{d}|c_{i}|^{2}=1,
\end{equation}
if and only if
\begin{equation}
\sum_{i=1}^{d}x_{i,j_{1}}^{*}x_{i,j_{2}}=0,
\end{equation}
for any $j_{1}\not=j_{2}$.

{\it Proof.} (If): To prepare $|\psi\rangle$, Alice performs a
transformation $Y=\sum_{i,j}y_{i,j}|i\rangle\langle j|$ on her
half in $|\Phi_{d}\rangle$ and then sends it to Bob's system. Thus
\begin{equation}
(Y\otimes I)|\Phi_{d}\rangle=|\psi\rangle,
\end{equation}
that is,
\begin{equation}
\sum_{i,j}^{d}c_{j}y_{i,j}|i\rangle_{A}|j\rangle_{B}=\frac{1}{\sqrt{d}}\sum_{i,j}^{d}x_{i,j}|i\rangle_{A}|j\rangle_{B}.
\end{equation}
Therefore,
\begin{equation} c_{j}y_{i,j}=\frac{x_{i,j}}{\sqrt{d}}\end{equation}
for any $i,j$. We know that the transformation $Y$ can be
successfully implemented with certain probability $P_{s}$ in terms
of Kraus operators $ E_{0}=\frac{Y}{\|Y\|}$,
$E_{1}=\sqrt{I-E_{0}^{\dag}E_{0}}$. Therefore,
\begin{equation}
P_{s}=Tr
E_{0}^{\dag}E_{0}\frac{I}{d}=\frac{Tr(Y^{\dag}Y)}{d\|Y^{\dag}\|\|Y\|}.
\end{equation}

Next, in order to show that $P_{s}$ can arrive at one, it suffices
to construct appropriate $y_{i,j}$ and $c_{j}$ such that $Y$ is
unitary and equations (18,22) hold. First we know that $Y$ is
unitary if and only if
\begin{equation}
\sum_{i=1}^{d}y_{i,j_{1}}^{*}y_{i,j_{2}}=\left\{\begin{array}{ll}
1,& j_{1}=j_{2},\\
0,& j_{1}\not=j_{2}.
\end{array}
\right.
\end{equation}
We take $c_{j}$ satisfying:
\begin{equation}
|c_{j}|^{2}=\frac{1}{d}\sum_{i=1}^{d}|x_{i,j}|^{2}, \hskip 2mm
j=1,2,\ldots,d.
\end{equation}
Clearly, if $|c_{j}|^{2}=0$, then $x_{i,j}=0$ for
$i=1,2,\ldots,d$. Furthermore we take $y_{i,j}$ in terms of the
following:
\begin{equation}
y_{i,j}=\left\{\begin{array}{ll} \frac{x_{i,j}}{\sqrt{d}c_{j}},&
{\rm if}\hskip 2mm c_{j}\not=0,\\
\frac{1}{\sqrt{d}},&{\rm otherwise}.
\end{array}\right.
\end{equation}
Now, in terms of equations  (19,25,26) and
$\langle\psi|\psi\rangle=1$, it is straightforward to check that
these $c_{j}$ and $y_{i,j}$ determined satisfy equations  (18,22).
The unitarity of $Y$ results in $Tr(Y^{\dag}Y)=d$ and
$\|Y^{\dag}\|=\|Y\|=1$. Thus, by equation (23) $P_{s}=1$ for the
constructed transformation $Y$.

(Only if): The known conditions say that there exists
transformation $Y=\sum_{i,j}y_{i,j}|i\rangle\langle j|$ such that
 equations  (18,22) hold and  $P_{s}=1$. From $P_{s}=1$ it follows that
\begin{equation}
Tr(Y^{\dag}Y)=d\|Y^{\dag}\|\|Y\|=d\|Y\|^{2}.
\end{equation}
Suppose that $\lambda_{i}\geq 0$ $(i=1,2,\ldots,d)$ are the
eigenvalues of $Y^{\dag}Y$. Then
$Tr(Y^{\dag}Y)=\sum_{i=1}^{d}\lambda_{i}$, and
$\|Y\|^{2}=\max(\lambda_{1},\lambda_{2},\ldots,\lambda_{d})$. If
there exist two different eigenvalues of $Y^{\dag}Y$, then
\begin{equation}
Tr(Y^{\dag}Y)=\sum_{i=1}^{d}\lambda_{i}<d\max(\lambda_{1},\lambda_{2},\ldots,\lambda_{d})=d\|Y^{\dag}\|\|Y\|.
\end{equation}
Consequently, $P_s<1$, a contradiction to $P_{s}=1$. Therefore, we
have $\lambda_{1}=\lambda_{2}=\ldots=\lambda_{d}=\lambda>0$ for
some $\lambda>0$. Thus, $Y^{\dag}Y=\sum_{i=1}^{d}\lambda
|e_{i}\rangle\langle e_{i}|$ for some orthonormal base
$\{|e_{i}\rangle\}$, which implies that
$\frac{Y^{\dag}Y}{\lambda}=I$ ($I$ denotes the identity operator).
This also shows that $\frac{Y}{\sqrt{\lambda}}$ is a unitary
operator. Therefore, for any $j_{1}\not=j_{2}$,
\begin{equation}
\sum_{i=1}^{d}y_{i,j_{1}}^{*}y_{i,j_{2}}=\langle
j_{1}|Y^{\dag}Y|j_{2}\rangle=0.
\end{equation}
From  equations (22,29) it follows directly that  equation (19)
holds. Therefore, we have completed the proof. \hfill $\Box$

{\it Remark.} In the above Theorem 1, state
$|\Phi_{d}\rangle=\sum_{i=1}^{d}c_{i}|i\rangle_{A}|i\rangle_{B}$
can be generalized to the more generic form
\begin{equation}
|\Phi_{d}\rangle=\sum_{i=1}^{d}c_{i}|g_i\rangle_{A}|h_i\rangle_{B}
\end{equation}
where
$\{|g_{i}\rangle_{A}:i=1,2,\ldots,d\}=\{|i\rangle_{A}:i=1,2,\ldots,d\}$,
and
$\{|h_{i}\rangle_{B}:i=1,2,\ldots,d\}=\{|i\rangle_{B}:i=1,2,\ldots,d\}$.
Therefore, there exist permutations $\Pi_{A}$ and $\Pi_{B}$ such
that $\Pi_{A}(i)=g_{i}$ and $\Pi_{B}(i)=h_{i}$ for
$i=1,2,\ldots,d$. The proof of the theorem with this change is
analogous, only by changing $Y$ to
$\sum_{i,j}y_{i,j}|g_i\rangle\langle g_j|$, by changing the left
side of equation (21) to
$\sum_{i,j}^{d}c_{j}y_{i,j}|g_i\rangle_{A}|h_j\rangle_{B}$, and in
places, by changing $x_{i,j}$ to $x_{\Pi_{A}(i),\Pi_{B}(j)}$.
\hfill $\Box$

 A straightforward corollary from Theorem 1 is as follows.

{\it Corollary 1.} Let
$|\psi\rangle=\frac{1}{\sqrt{d}}\sum_{i,j=1}^{d}x_{i,j}|i\rangle|j\rangle$
be the state to be prepared. If equation (19) holds, i.e., there
exist $j_{1}\not=j_{2}$ such that $
\sum_{i=1}^{d}x_{i,j_{1}}^{*}x_{i,j_{2}}\not=0, $ then for any
state $|\Phi\rangle=\sum_{i=1}^{d}c_{i}|i\rangle_{A}|i\rangle_{B}$
with $\sum_{i=1}^{d}|c_{i}|^{2}=1$, initially shared by Alice and
Bob, the success probability for preparing $|\psi\rangle$ is
strictly smaller than one.

Naturally, we may ask how about the lower bound on the success
probability for superdense coding if the condition described by
equation (19) does not hold. Next we consider the case of which
equation (19) does not hold only for arbitrarily given two $j_{1}$
and $j_{2}$, and for the others, equation (19) is still preserved.

{\it Proposition 2.} Let
$|\psi\rangle=\frac{1}{\sqrt{d}}\sum_{i,j=1}^{d}x_{i,j}|i\rangle|j\rangle$
be the state to be prepared. If for $\{j_{1},j_{2}\}=
\{k_{1},k_{2}\}$, equation (19) does not hold, but for the other
cases, equation (19) is preserved, then by initially sharing
$|\Phi\rangle=\sum_{i=1}^{d}c_{i}|i\rangle_{A}|i\rangle_{B}$
between Alice and Bob, where $\sum_{i=1}^{d}|c_{i}|^{2}=1$, the
maximum success probability $P_{s}^{(m)}$ for preparing
$|\psi\rangle$ satisfies
\begin{equation}
1>P_{s}^{(m)}\geq
\left(1+\frac{1}{d|c_{k_{1}}c_{k_{2}}|}\left|\sum_{i=1}^{d}x_{i,k_{1}}^{*}x_{i,k_{2}}\right|\right)^{-1}.
\end{equation}

{\it Proof.} First, $P_{s}^{(m)}<1$ follows directly from Theorem
1. Next we prove the other inequality. We take $c_{j}$ as equation
(25), i.e., $|c_{j}|^{2}=\frac{1}{d}\sum_{i=1}^{d}|x_{i,j}|^{2}$.
As above, let $Y=\sum_{i,j=1}^{d}y_{i,j}|i\rangle\langle i|$ be
the transformation on system $A$ performed by Alice. Then equation
(22) holds, i.e., for any $i,j$,
$c_{j}y_{i,j}=\frac{x_{i,j}}{\sqrt{d}}$. By taking
\[
y_{i,j}=\left\{\begin{array}{ll} \frac{x_{i,j}}{\sqrt{d}c_{j}},&
{\rm if}\hskip 2mm c_{j}\not=0,\\
\frac{1}{\sqrt{d}},&{\rm otherwise},
\end{array}\right.
\]
we then have $Tr(Y^{\dag}Y)=d$, and
\begin{eqnarray*}
Y^{\dag}Y&=&\sum_{i,j_{1},j_{2}=1}^{d}y_{i,j_{1}}^{*}y_{i,j_{2}}|j_{1}\rangle\langle
j_{2}|\\
&=&\sum_{j=1}^{d}|j\rangle\langle
j|+\sum_{i=1}^{d}\frac{x_{i,k_{1}}^{*}x_{i,k_{2}}}{dc_{k_{1}}^{*}c_{k_{2}}}|k_{1}\rangle\langle
k_{2}|+\sum_{i=1}^{d}\frac{x_{i,k_{2}}^{*}x_{i,k_{1}}}{dc_{k_{2}}^{*}c_{k_{1}}}|k_{2}\rangle\langle
k_{1}|.
\end{eqnarray*}
Then we can determine that the eigenvalues of $Y^{\dag}Y$ are 1,
and $1\pm
\frac{1}{d|c_{k_{1}}c_{k_{2}}|}\left|\sum_{i=1}^{d}x_{i,k_{1}}^{*}x_{i,k_{2}}\right|$.
Therefore, by virtue of equation (23) we obtain $
P_{s}^{(m)}\geq\frac{Tr(Y^{\dag}Y)}{d\|Y^{\dag}\|\|Y\|}=
\left(1+\frac{1}{d|c_{k_{1}}c_{k_{2}}|}\left|\sum_{i=1}^{d}x_{i,k_{1}}^{*}x_{i,k_{2}}\right|\right)^{-1}$,
the lower bound as desired. \hfill $\Box$

Especially, if $\sum_{i=1}^{d}x_{i,k_{1}}^{*}x_{i,k_{2}}=0$, then
the above bound described by inequality (31) reduces to $1$,
complying with Theorem 1.

\section*{4. Concluding Remarks}

Superdense coding of quantum states, first proposed by Harrow,
Hayden, and Leung [16], describes that if the sender knows the
identity of the state to be sent, then two qubits can be
communicated with a certain probability by physically transmitting
one qubit and consuming one bit of entanglement. The objective of
this protocol is to prepare a quantum state in Bob's system or
``sharing" a state that is entangled between Alice and Bob's
systems, for which Alice and Bob initially share a (maximally)
entangled state, and Alice first performs a physically operation
on her party with a certain success probability and then sends it
to Bob. Furthermore, Harrow {\it et al.} [18] presented a protocol
succeeding with high probability for communicating a $2l$-qubit
quantum state but some shared random bits are necessarily
consumed, besides transmitting $l+o(l)$ qubits and consuming $l$
ebits of entanglement. Notably, if the shared randomness is {\it
not} required, Hayden, Leung and Winter [20] proposed a different
protocol of superdense coding of quantum states that can always
successfully perform the physical process, but may not guarantee
the result to be {\it exact}. Rather, the protocol may result in
an approximate outcome with high fidelity.

In this paper, we discovered that some states are impossible to be
perfectly prepared if Alice and Bob initially share the entangled
states that are superposed by the ground states, as the states to
be prepared. Particularly, we gave a sufficient and necessary
condition for the states being enabled to be exactly prepared with
probability one, by initially sharing these entangled states
(maybe {\it not maximally}) between Alice and Bob. Furthermore,
the lower bound on the probability for preparing some states was
derived. Thus, this is another profile regarding superdense coding
of quantum states. Also, in a way, this partially makes up the
existing outcomes [18,20].

As well, for exactly preparing some quantum states, we determined
some partially entangled states initially shared by Alice and Bob
that result in the optimal success probability one. However, if,
instead, the initial entanglement shared by the two parties is
maximal, then the success probabilities for preparing these states
may be smaller than one, a different phenomenon from superdense
coding [3,12,14].

\section*{Acknowledgement}
This work is supported by the National Natural Science Foundation
(No. 90303024, 60573006), the Research Foundation for the Doctoral
Program of Higher School of Ministry of Education (No.
20050558015), and the Natural Science Foundation of Guangdong
Province (No. 031541) of China.


\begin{thebibliography}{AB}
\bibitem{ab}Nielsen M A and Chuang I L 2000 {\it Quantum Computation and Quantum Information}
  (Cambridge: Cambridge University Press)


\bibitem{ab} {\it Quantum Information: An Introduction to
Basic Theoretical Concepts and Experiments (Springer Tracts in
Modern Physics; 173)}, edited by G. Albert, T. Beth, M. Horodecki,
{\it et al.} 2001 (Berlin: Springer)


\bibitem{ab}Bennett C H and  Wiesner S J 1992 {\it Phys. Rev. Lett.} {\bf 69} 2881

\bibitem{ab}Bennett C H, Brassard G, Crepeau C, Jozsa R, Peres  A and  Wootters W K 1993 {\it Phys. Rev. Lett.} {\bf 70} 1895

\bibitem{ab} Lo H-K 2000 {\it Phys. Rev. A} {\bf 62}
012313;\\ Bennett C H, Hayden P,  Leung D,  Shor P W and Winter A
2005 {\it IEEE Trans. Inf. Theory} {\bf 51} 56

\bibitem{ab} Shor P W 1997  {\it SIAM J. Comp.} {\bf 26} 1484


\bibitem{ab}Gisin N, Ribordy G,  Tittel W and  Zbinden H, {\it Rev. Mod.
Phys.} {\bf 74} 145

\bibitem{ab} Werner R F 2001 {\it J. Phys. A} {\bf 34} 7081; \\
Hao J-C,  Li C-F and  Guo G-C 2000, {\it Phys. Lett. A} {\bf 278}
113


\bibitem{ab} Wu S,  Cohen S M,  Sun Y and Griffiths R B 2006 {\it Phys. Rev. A} {\bf 73}
042311



\bibitem{ab} Bose S,  Vedral V and Knight P L 1998 {\it Phys. Rev. A} {\bf 57}
822; \\X.  Liu S,  Long G L,  Tong D M and  Li F 2002 {\it Phys.
Rev. A} {\bf 65} 022304; \\ Bru$\ss$ D,  Lewenstein M, Sen(De) A,
 Sen U,  D'Ariano G M and  Macchiavello C, quant-ph/0507146

\bibitem{ab} Barenco A and  Ekert A 1995 {\it J. Mod. Opt.} {\bf 42} 1253

\bibitem{ab} Hausladen P,  Jozsa R,  Schumacher B,  Westmoreland M and  Wootters W K 1996 {\it Phys. Rev. A} {\bf 54}
1869

\bibitem{ab} Bowen G 2001 {\it Phys. Rev. A } {\bf 63}
022302
\bibitem{ab} Mozes S,  Reznik B and  Oppenheim J 2005 {\it Phys. Rev. A} {\bf 71}
012311


\bibitem{ab} Pati A K,  Parashar P and   Agrawal P 2005 {\it Phys. Rev. A} {\bf 72}
012329


\bibitem{ab} D\"{u}r W, Vidal G and Cirac J I 2000  {\it Phys. Rev. A} {\bf 62}
062314



\bibitem{ab}  Agrawal P and  Pati A K,  quant-ph/0610001, {\it Phys. Rev.
A} (to be published)


\bibitem{ab} Harrow A, Hayden P and  Leung D W 2004 {\it Phys. Rev. Lett.} {\bf 92} 187901


\bibitem{ab} Abeyesinghe A,  Hayden P and  Smith G
2006 {\it IEEE Trans. Inf. Theory} {\bf 52} 3635


\bibitem{ab}  Hayden P, Leung D W  and   Winter A 2006 {\it Commun. Math. Phys.}  {\bf 265} 95


\bibitem{ab} Kraus K 1983 {\it States, Effects, and Operations}
(Berlin: Springer-Verlag)
\bibitem{ab} Peres A 1995 {\it Quantum Theory: Concepts and Methods}
(Dordrecht: Kluwer)












\end{thebibliography}
\end{document}